# Surface Reactivity in Low Temperature Deposited Amorphous–Crystalline SnO₂ Thin Films: Chemisorbed Oxygen Activity and CO Oxidation Pathways Revealed by In Situ XPS and Mass Spectrometry

Engin Ciftyurek[1], Zheshen Li[2] and Klaus Schierbaum[1]

[1]Department of Materials Science, Institute for Experimental Condensed Matter Physics, Heinrich Heine University of Düsseldorf, 40225 Düsseldorf, Germany

[2]ISA, Centre for Storage Ring Facilities, Department of Physics and Astronomy, Aarhus University, Ny Munkegade 120, 8000C Aarhus, Denmark

## Abstract

This study investigates two critical aspects of the gas sensing mechanism in metal oxide sensors: (1) the conditions that maximize chemisorbed oxygen concentration as a function of temperature and oxygen partial pressure, and (2) which surface oxygen species—chemisorbed or lattice-bound—are primarily responsible for interaction with carbon monoxide (CO). $SnO_2$ thin films, deposited at temperatures as low as 60 °C and exhibiting mixed amorphous-crystalline phases with open, tortuous porosity, were evaluated for CO sensing at 200 °C. Comprehensive characterization using EIS, MS, XPS, TEM, and sensor tests revealed a strong correlation between high sensing performance and the structural/electronic features of the defect rich low-temperature-deposited $SnO_2$. Electrochemical impedance spectroscopy (EIS) was employed to identify the optimal sensing temperature. Mass spectroscopy (MS) used to analyze the exhaust gases after sensing reactions. The films exhibited oxygen under-stoichiometry and high concentrations of chemisorbed oxygen species. In situ XPS under 1 mbar (10000 ppm) $O_2$ and CO exposures showed that chemisorbed oxygen, not lattice oxygen, was actively involved in CO oxidation, as further confirmed by $CO_2$ detection via Mass spectroscopy (MS). Quantitative analysis revealed dynamic surface chemical status alternations, emphasizing the pivotal role of chemisorbed oxygen in the sensing mechanism at 200°C.

## 1. Introduction

Tin dioxide ($SnO_2$) is one of the most extensively studied n-type semiconducting metal oxides (SMOs) for gas sensing applications due to its low cost, high thermal and chemical stability, and robust sensitivity to a range of reducing and oxidizing gases. $SnO_2$-based sensors are particularly valued for detecting toxic gases such as carbon monoxide (CO), which is colorless, odorless, and detrimental even at low concentrations. These sensors typically operate on the principle of resistance modulation driven by redox reactions at the metal oxide surface. Despite decades of research, the mechanistic and physico-chemical details underlying this sensing response—especially in low-temperature regimes—remain a topic of active investigation. Disentangling the roles of chemisorbed and lattice oxygen remains a critical challenge, as it directly impacts the interpretation of sensing behavior, sensor design strategies, and performance optimization (1) (2) (3; 4; 5; 6; 7; 8; 9).

A critical aspect of sensing mechanism involves elucidating the electronic response of $SnO_2$ in the presence of background oxygen. Historically, the interaction between oxygen and $SnO_2$ has been interpreted through the lens of the "ionosorption theory," wherein adsorbed oxygen species are considered free oxygen ions that are electrostatically stabilized on the surface without forming local chemical bonds. For the specific example of $SnO_2$ surfaces adsorb oxygen molecules that trap conduction band electrons, forming negatively charged chemisorbed species such as $O_2^-$ and $O^-$, when exposed to a reducing gas such as CO, these chemisorbed oxygen species are consumed, resulting in a decrease in electrical resistance. These species are believed to serve as the active sites for redox reactions with reducing gases.

A few recent studies challenge the conventional ionosorption framework by presenting evidence for the direct participation of lattice oxygen in redox reactions with reducing gases such as $SO_2$, CO, $H_2$, and $H_2S$. This apparent contradiction highlights the need to re-examine the respective contributions of chemisorbed and lattice oxygen species under varying temperatures and oxygen partial pressures. Notably, in situ analyses combined with theoretical calculations at ~200 °C suggest that lattice oxygen can replenish the chemisorbed oxygen pool and actively contribute to the sensing response, thereby questioning the strictly ionosorption-based interpretation (10). However, this low-temperature perspective stands in contrast to the more established view that significant reduction of the oxygen sublattice in $SnO_2$ is rarely observed at ~200 °C, even under oxygen-deficient conditions. Instead, the prevailing consensus is that at higher temperatures (>500 °C), oxygen adsorption becomes stabilized primarily through its incorporation into the lattice as $O^{2-}$ ions at vacancy sites. The dynamic equilibrium of these lattice-stabilized species with the ambient oxygen partial pressure is believed to underpin sensor functionality at elevated temperatures, providing a more robust explanation for long-term stability and response (3). Another study support this dynamic behavior is a computational (DFT) study that models CO oxidation occurring via Mars–van Krevelen mechanism on the $SnO_2$(110) surface, where lattice oxygen is directly involved in oxidizing CO and being replenished by gas-phase $O_2$ (11).

For high temperature sensor operations (>500°C) the possible involvement of lattice oxygen in these surface reactions has also been anticipated due to experimental evidence. While chemisorbed oxygen is known to remain on the surface of metal oxides up to temperatures of approximately >400 °C, certain doped variants of $SnO_2$, such as $Gd_{1.8}Y_{0.2}Zr_2O_7$ and $Gd_{1.6}Sm_{0.4}Zr_{1.9}Sn_{0.1}O_7$, remain functional at temperatures exceeding 500 °C. This raises important questions about the thermal stability of the oxide lattice and the potential involvement of lattice oxygen in surface redox reactions during sensing (12; 13; 14; 15; 16).

Two central questions are fundamental to the issue indicated above: (1) under what thermodynamic and kinetic conditions is the concentration of chemisorbed oxygen species maximized, and (2) which specific oxygen species—chemisorbed or lattice-bound—are primarily responsible for the redox interaction with target gases. The first of these questions has been addressed in our previous studies through a combination of adsorption and sensor studies, electrical characterization, and surface-sensitive techniques for chemical analysis such as XPS, XPEEM, NAP-XPS and UPS (3; 5; 17; 9). These studies have established that chemisorbed oxygen concentration depends strongly on surface stoichiometry, temperature, oxygen partial pressure, and the presence of surface defects such as oxygen vacancies (18; 19; 20; 2).

The second question—the identity of the reactive oxygen species during gas sensing—remains less well understood. This work aims to address that question by combining operando mass spectrometry, in situ XPS analysis, and resistive gas sensing under controlled CO and $O_2$ exposures. Our experimental strategy focuses on correlating the evolution of chemisorbed oxygen species with the dynamic sensor response and $CO_2$ production under low-temperature (200°C) CO sensing conditions. Through this approach, we aim to clarify whether chemisorbed oxygen ions alone account for the sensor signal, or whether lattice oxygen also participates in the surface redox processes. The outcome of this study is expected to provide a more accurate understanding of the $SnO_2$–CO interaction mechanism and inform the rational design of next-generation, low-temperature metal oxide gas sensors with enhanced selectivity and efficiency.

## 2. Application of In-Situ XPS in Studying Surface Chemistry of Functional Materials for Gas Sensing

XPS is a powerful technique for probing the chemical composition of surfaces under realistic reaction conditions (21; 22; 23). Recent advances in in-situ, operando, and near-ambient pressure (NAP) XPS have greatly expanded our understanding of surface chemistry in metal oxides and related materials across diverse applications, including sensors, catalysis, and solid oxide fuel cells (24). In-Situ XPS studies on $TiO_2$, $WO_3$, $LaMnO_3$, $LaCoO_3$, and $WS_2$-based systems have revealed critical insights into surface interaction mechanisms in 2D semiconductors, thin films, and perovskite materials under reactive CO and $O_2$ environments. Similarly, in-situ XPS of $TiO_2$ under varying pressures of ethanol adsorption demonstrated predominantly physisorptive behavior, with only limited chemisorbed ethoxide formation (25; 26; 27; 28; 29). These methods enable quantification of dynamic oxygen species before and after CO exposure in catalytic and sensing applications involving complex oxides and layered materials (7; 1; 30).

In-Situ XPS has also been applied to nanoparticle catalysts such as Pt–Rh, Pd–Rh, and Pd–Pt (31; 32) (33). In ceria-based fuel cells, operando XPS identified reaction intermediates via O 1s analysis, elucidating pathways of $CO_2$ hydrogenation and ethylene oxidation (34) (35). Likewise, Zhang and Chung et al. reported the role of chemisorbed oxygen on Pt oxides under varying $O_2$/CO ratios (36; 37). A few reports in recent years highlighted the vital role of the surface chemical properties under real sensor testing conditions. Ciftyurek at. al employed In-Situ XPS in NAP-XPS configuration for the first time to analyze the metal oxide interaction with reducing gases. Moreover, combined synchrotron-based UPS/XPS and NAP-XPS studies on $WO_3$ thin films have provided temperature-resolved quantification of chemisorbed oxygen species during exposure to reactive gas mixtures, underlining the utility of these techniques in operando analysis (22; 6; 5; 38; 39) (40). The accepted mechanism of a chemical sensor involves the interaction of negatively charged oxygen ($O^{2-}, O_2^{2-}, O_2^-, O^-$) and target gases. The In-situ XPS will lead to understand that CO interaction with $SnO_2$ surface lattice component and adsorbed oxygen ions. In the current work, we studied interaction between chemisorbed oxygen with CO using XPS under realistic pressures up to 10000 ppm (1 mbar).

## 3. Low Temperature (60°C) PE-ALD Deposited 20 nm $SnO_2$

### 3.1.  TEM Transmission Electron Microscopy (TEM)

Figure 1-a and 1-b shows cross-section TEM image of 20 nm $SnO_2$ deposited at 60°C for microstructural and crystallographic analysis. The sharp interfaces between Si, $SiO_2$, and $SnO_2$ suggest good control of deposition and absence of interdiffusion.

The crystallinity gradient from interface to surface suggests SnO$_2$ growth begins as nanocrystalline islands, evolving toward more defined grains away from SiO$_2$. The pore free, homogeneous thin film with structural integrity and continuity are visible in TEM images. The SnO$_2$ layer shows fine lattice fringes—these are indicative of crystallinity but not perfectly uniform, suggesting a nanocrystalline or mixed amorphous-crystalline structure. Lattice fringes of ~0.33 nm for SnO$_2$ (110) plane was observed. The ~2 nm in thickness SiO$_2$ interface between the Si substrate and the SnO$_2$ is observed. This SiO$_2$ formation is due to the applied PE-ALD deposition parameter with relatively long plasma pulse times of 500 ms and rapid formation nature of SiO$_2$ native oxide on Si surface. Presence of SiO$_2$ interlayer ensures electrical isolation and prevents silicide formation at elevated sensor testing temperatures.

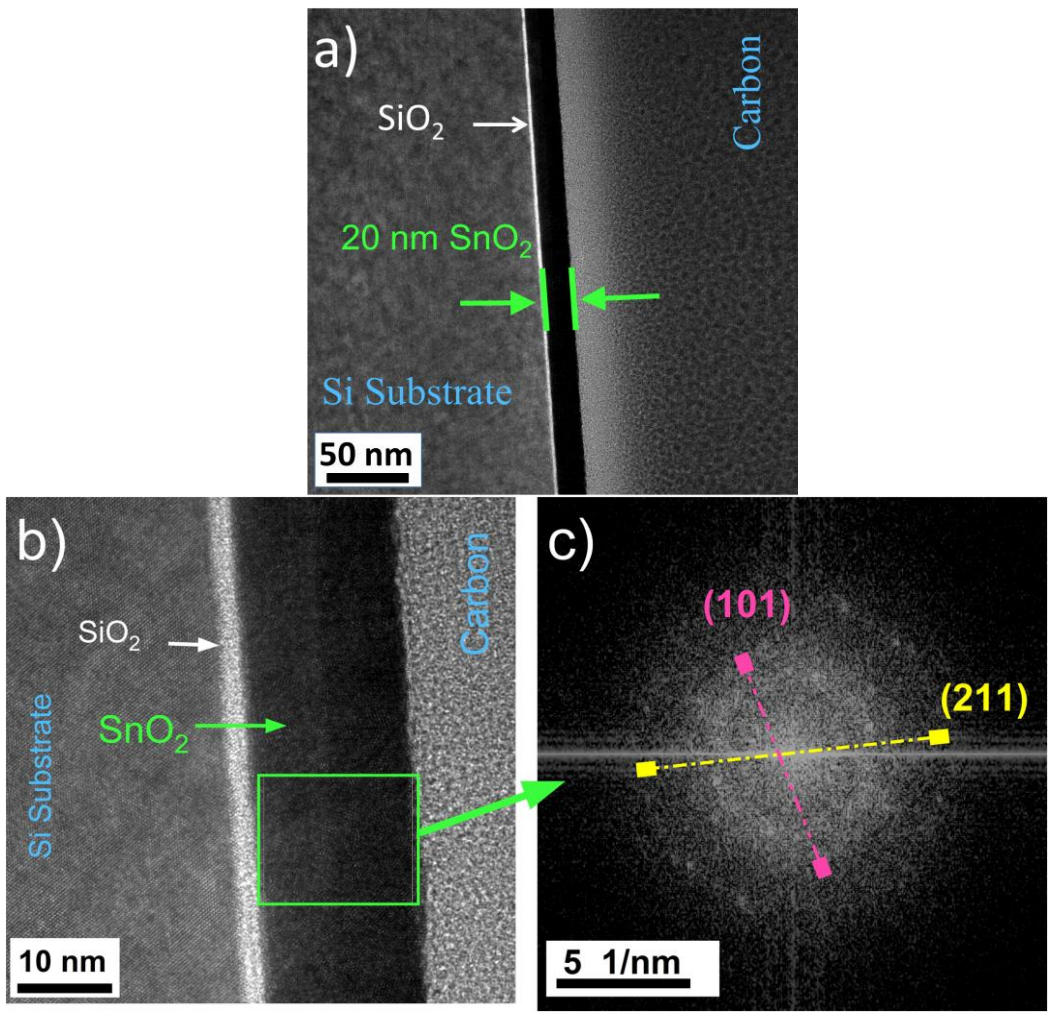

**Figure 1: (a-b)** Cross-sectional images showing the Si/SnO$_2$/Carbon with SiO$_2$ native oxide **(c)** FFT diffractogram obtained from the image with green selected area of the SnO$_2$ layer.

In order to confirm the phase of the SnO$_2$ later we performed selected area diffraction pattern analysis. Figure 1-c presents the Fast Fourier transformation (FFT) diffraction pattern of the green square region. FFT identified only two crystal planes of (101) and (211) confirming the nanocrystallinity of SnO$_2$ with polycrystalline nature, while diffuse rings point out the existence of amorphous regions (41). The TEM and FFT analysis indicate that the SnO$_2$ film is defect-rich, as evidenced by the presence of diffuse diffraction rings rather than sharp, well-defined spots, which suggests the material is nanocrystalline with very small grain sizes. The broadening and intensity variations of the diffraction features can be attributed to oxygen vacancies and grain boundary effects that disrupt the long-range crystallinity.

All investigated SnO₂ thin films exhibit a polycrystalline nature, characterized by weak yet distinct diffraction peaks corresponding to the cassiterite tetragonal phase. The under developed crystallinity is consistent with prior findings (42; 43; 44), particularly for metal oxides grown at low temperatures, where grazing incidence X-ray diffraction (GI-XRD) at 60 °C revealed a coexistence of amorphous and nanocrystalline domains. Such structural disorder is common in thin-film SnO₂ synthesized under non-equilibrium conditions (low temperature) and implies a high density of grain boundaries, oxygen vacancy point defects and open porosity (beneficial for gas sensing) (45).

While these defects can reduce carrier mobility and optical transparency in electronic or optoelectronic applications, they are beneficial for gas sensing, as oxygen vacancies and surface defects enhance surface reactivity and adsorption processes, thereby improving sensitivity and response characteristics. This structural heterogeneity—marked by partially developed crystallinity and the presence of disordered amorphous regions—facilitates a high density of surface defects and under-coordinated sites. The defect-rich and mixed-phase morphology are contributing to the accumulation of chemisorbed oxygen species, which play a pivotal role in enhancing gas sensing performance, particularly in redox reactions involving carbon monoxide.

## 4. Investigation of SnO₂ Surface Chemistry under CO and O₂ Exposures: Insight into Gas-Solid Interactions

Tin dioxide (SnO₂) remains one of the most widely studied semiconducting metal oxides for chemical gas sensors, particularly due to its high sensitivity to reducing gases. Despite decades of research, the sensing mechanism—especially the interaction between SnO₂ and target gases like carbon monoxide (CO)—is still not fully understood. A central topic of debate concerns the roles of different oxygen species, particularly chemisorbed oxygen versus lattice oxygen, in governing the surface redox reactions that drive sensor response.

### 4.1. Oxygen interaction with the SnO₂ Surface: Chemisorbed Oxygen Ions Dominate the Sensing Mechanism

The chemisorption of molecular oxygen ($O_2$) on SnO₂ surfaces primarily proceeds via dissociative adsorption, a process that is energetically favorable only at elevated temperatures due to the substantial activation energy barrier (~5 eV) (46; 47). This mechanism is strongly influenced by the presence of surface oxygen vacancies, which facilitate the formation of reactive chemisorbed oxygen species. Depending on the specific metal oxide, surface stoichiometry, and vacancy concentration, this dissociative adsorption becomes particularly efficient in the temperature range of 150–250°C.

As shown in Figure 2-a, XPS analysis performed at 25 °C prior to O₂ exposure revealed three distinct oxygen species in the O 1s core-level spectrum: lattice oxygen ($O^{2-}$) at 530.8 eV, chemisorbed oxygen ($O^{-}$) at 531.8 eV, and hydroxyl/water-related species ($OH^{-}/H_2O$) at 532.7 eV. At this stage, the chemisorbed oxygen concentration was ~13.5 at.%, a relatively high value attributed to the low-temperature deposition process that produced a partially amorphous, porous SnO₂ structure rich in disorder and oxygen vacancies.

As shown in Figure 2-b, upon heating to 200°C and exposure to 10.000 ppm (1 mbar) $O_2$, the binding energies of the oxygen species (lattice oxygen, chemisorbed oxygen and hydroxyl/water-related species) remained unchanged. However, the amount of surface-adsorbed water ($OH^-/H_2O$) terminated, while the chemisorbed oxygen ion ($O^-$) concentration significantly increased to ~22.5 at.%. This rise reflects enhanced $O_2$ dissociation, promoted by the abundance of vacancy-rich sites on $SnO_2$ surface. These quantitative shifts in surface oxygen species—summarized in Table 1—emphasize the decisive role of surface characteristics (porosity and oxygen-vacancy defects) in governing the gas-sensing mechanism of $SnO_2$.

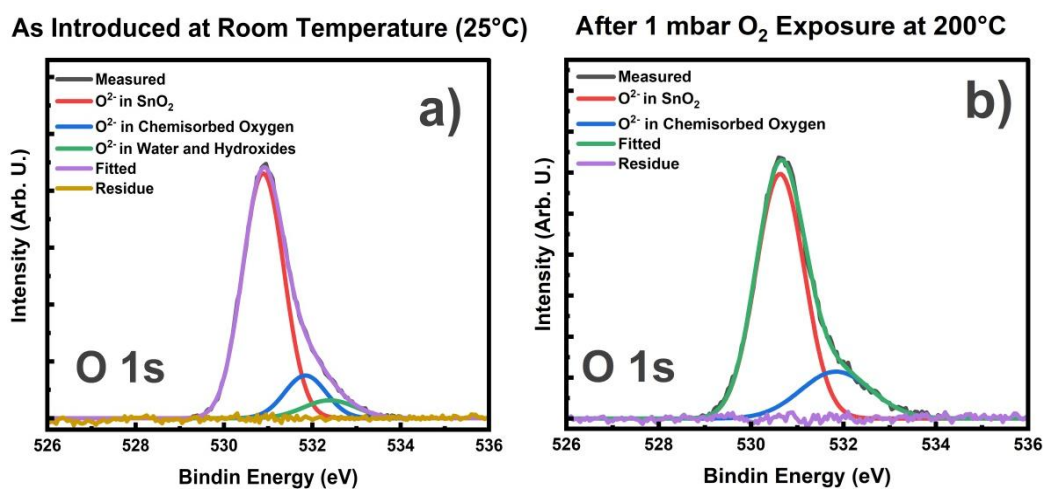

**Figure 2:** O 1s XPS core-level spectra showing **(a)** the presence of chemisorbed oxygen at 25 °C and **(b)** a substantial increase in chemisorbed oxygen after exposure to 10,000 ppm $O_2$ (1 mbar) at 200 °C.

## 4.2. CO Exposure of SnO$_2$ Enriched with Chemisorbed Oxygen

Upon exposure to 5 min. 10000 ppm (1 mbar) of CO, a major change in the O 1s spectral profile was observed, as shown in Figure 3. The peak associated with chemisorbed oxygen ($O^-$) decreased remarkably, indicating their consumption in redox reactions with CO, also reflected in quantitative compositional data in Table 1.

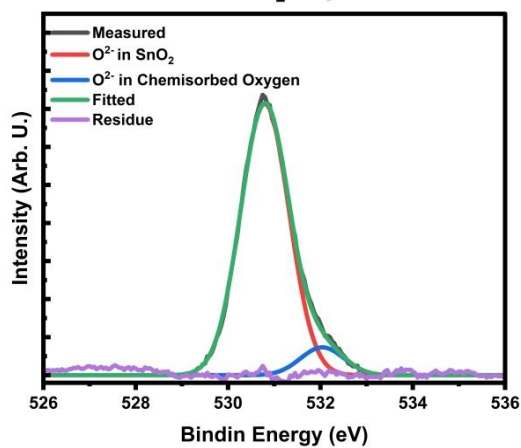

**Figure 3:** XPS O 1s core-level spectra after exposure to 10,000 ppm (1 mbar) CO. Following CO exposure, the concentration of chemisorbed oxygen ions ($O^-$) on the $SnO_2$ surface decreased to ~8.4 at.% (see Table 1).

While several intermediate ionic forms of oxygen species may be present at room temperature as chemisorbed oxygen species, among them the super oxide ion ($O_2^-$) is stable compared to the gaseous $O_2$. As temperature increases, diatomic oxygen ion ($O_2^-$) can transform into $O^-$ or desorb from the metal oxide surface. In the 150–200°C range, $O^-$ is the dominant reactive oxygen ion involved in sensing reactions in the current work.

The Eqs. 1 and 2 summarize the key redox reactions and explain the formation of $O^-$ on the surface and how CO consumes chemisorbed oxygen ions ($O^-$), leading to sensor response. Table 1 shows the observed evolution of surface oxygen species under CO exposure supports a mechanism dominated by the consumption–replenishment cycle of chemisorbed oxygen ions ($O^-$), providing critical insight for the rational design of high-sensitivity metal oxide gas sensors.

$$O_{2(gas)} \rightarrow O_{2\ (adsorbed)} \rightarrow O_2^-{}_{(chemisorbed)} \rightarrow O^-{}_{(chemisorbed)} \quad \textbf{(Eq. 1)}$$
$$CO + O^-{}_{(chemisorbed)} \rightarrow CO_{2\ (gas)} + e^- \quad \textbf{(Eq. 2)}$$

**Table 1.** Concentration of different Oxygen Species at 25°C and upon exposures to $O_2$ and CO at 200°C. Experimental evidence for oxygen species alteration upon surface redox reactions with CO.

| Oxygen ion concentration [at.%.] upon $O_2$ and CO Exposures | Oxygen ion in Water Hydroxides ($H_2O/OH^-$) | Oxygen ion in Chemisorbed Oxygen ($O_{2(ads)}^-, O_{(ads)}^-, O_{2(ads)}^{2-}$) | Oxygen ion in $SnO_2$ Lattice |
|---|---|---|---|
| Under Ultra High Vacuum (UHV) at 25°C | 7 | 13.5 | 79.5 |
| After 10000 ppm (1 mbar) $O_2$ at 200°C | < 0.1 | 22.5 | 77.5 |
| After 10000 ppm CO (1 mbar) at 200°C | < 0.1 | 8.4 | 91.6 |

## 4.3. Sn 3d Core Level Analysis

Figure 4 shows the Sn (3d) core level spectrums of as introduced $SnO_2$ sensor at 25°C and after sensor exposed to CO at 200°C. The fwhm values for Sn ($3d_{5/2}$) and Sn ($3d_{3/2}$) were 0.7 eV before and after exposure to CO at 200°C. The spin orbit splitting values of Sn (3d) were 8.5 eV (48). The observed values of both fwhm and spin-orbit splitting in the core Sn (3d) did not change after exposure to CO at 200°C.

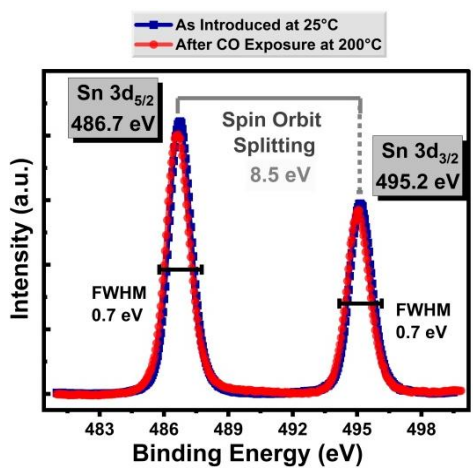

**Figure 4:** Sn 3d core-electron peaks do not show shift in electron binding energy after CO exposure, indicating that lattice oxygen ions in $SnO_2$ do not involve in the chemical sensing reactions.

We identified that $Sn^{4+}$ found in $SnO_2$ is the major chemical state. Binding Energy for Sn $3d_{5/2}$ and Sn $3d_{3/2}$ is measured to be at 486.7 eV and 495.2 eV matching with the $Sn^{4+}$ oxidation state (13; 49; 42); shift to lower energy after CO exposure ($Sn^{4+} \rightarrow Sn^{2+}$) was not observed so CO was seen to be just interacting with chemisorbed oxygen ions at 200°C. O 1s spectra revealed changes in chemisorbed oxygen ions ($O^-$) upon exposure to CO, while Sn 3d spectra did not show shifts in binding energy consistent with surface redox reactions. These data confirm active participation of chemisorbed oxygen species during CO sensing.

## 5. Impedance Spectroscopy (IS) analysis of $SnO_2$

The IS result collected over from 200°C to 250°C for $SnO_2$ sensor is presented in Figure 5. The figure shows impedance spectroscopy measurements of a $SnO_2$ sensor layer in air using a Bode–Nyquist plot at operating temperatures between 200°C and 250°C. Each color-coded curve corresponds to a different temperature and demonstrates how the sensor's impedance changes with temperature.

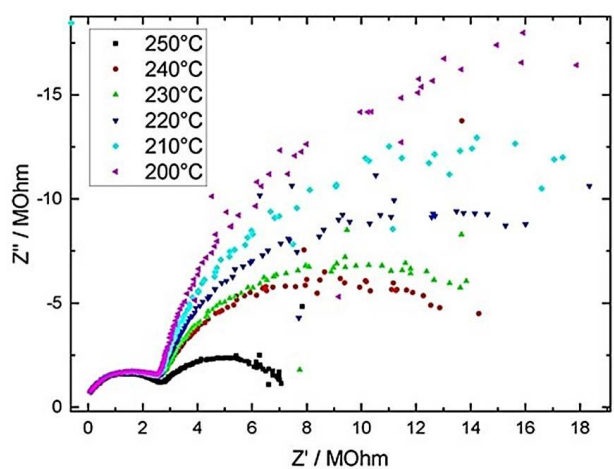

**Figure 5:** Impedance spectroscopy measurements of the $SnO_2$ shown in Nyquist representations.

By applying an alternating current (AC) signal over a range of frequencies, it possible to distinguish the effects of different elements to the total resistance (impedance). First semicircle (high-frequency) represents bulk resistance ($R_b$) and conduction through $SnO_2$ grains. Second semicircle (low-frequency) is associated with grain boundary resistance ($R_{gb}$) due to potential barriers at grain boundaries.

The Nyquist plots revealed an ideal semicircular shape in the high-frequency region, which remains completely unaffected by temperature variations. This consistent high-frequency region is attributed to the platinum (Pt) electrode and bulk resistance thus not related to the sensing capability. The low-frequency region exhibits a significant change with increasing temperature, a typical characteristic of polycrystalline semiconducting materials such as $SnO_2$. Specifically, at 200°C, the contribution from the $SnO_2$ layer in the low-frequency domain is highest in magnitude in comparison to the other temperatures tested emphasizing that the most sensitive sensor's gas detection capabilities will be obtained at 200°C.

As the temperature gradually increases from 200°C to 250°C, the semi-circles shift toward lower real impedance values; indicating improved electron conductivity and faster surface reaction kinetics. The observed behavior is primarily linked to the depletion of chemisorbed oxygen species on the sensor surface, thus release of electrons back to the conduction band. As previously reported by current authors, the stability of chemisorbed oxygen ions on metal oxide surfaces depends strongly on temperature, which varies for different metal oxides. In the case of oxygen vacancy rich SnO₂, the data show that around 200°C marks the critical point where surface-stabilized chemisorbed oxygen ions (O⁻) begin to transition toward desorption. As the temperature increases toward 250°C, this desorption process dominates, drastically reducing the concentration of chemisorbed oxygen ions (O⁻) on the surface (50; 17).

## 6. SnO₂ Testing for CO; Sensor Response and Insights

Figure 6 illustrates the dynamic response of the SnO₂-based gas sensor to CO exposure at 200 °C. The interaction between CO molecules and pre-adsorbed chemisorbed oxygen species initiates surface oxidation, yielding CO₂ and releasing electrons into the conduction band of SnO₂. This electron donation reduces the sensor's electrical resistance, serving as the primary detection signal. The observed high resistance values (in the MΩ range) before and after CO exposure indicate the active participation of the amorphous component in the low-temperature (60 °C) PEALD-deposited SnO₂ thin films (8).

The sensor response continues until chemisorbed oxygen ions around the grain boundaries—responsible for maintaining Schottky barriers—is sufficiently depleted, resulting in signal saturation. Notably, this indicates that the sensing response is governed solely by the removal of chemisorbed oxygen, with no evidence of lattice oxygen involvement. As detailed in Table 1, chemisorbed oxygen species decreased from 22.5 at.% to 8.4 at.% after CO exposure, in parallel to this with a high sensor response of 32 for 135 ppm of CO was observed at 200°C; both go in line with the mass spec data presented in the nest section.

Sensor Signal under CO exposures, demonstrating fast, reversible response behavior. Decreases in resistance coincide with CO-induced reduction of chemisorbed oxygen. Proposed mechanism of CO interaction with chemisorbed oxygen species on the SnO₂ surface at 200 °C based on the XPS, Impedance spectroscopy and dynamic sensor testing. The adsorption of oxygen, formation of chemisorbed species (O⁻), and their subsequent reaction with carbon monoxide (CO) to produce carbon dioxide (CO₂) and releasing conduction electrons back to SnO₂ conduction band, decreasing the resistance of the sensor, finally regeneration of the surface via background O₂.

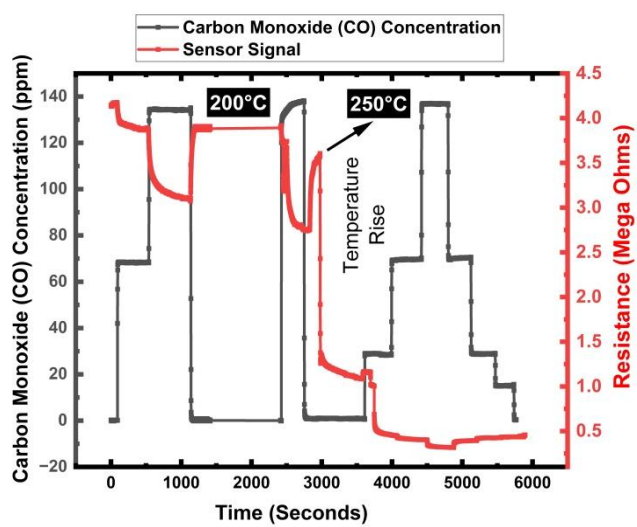

**Figure 6:** Dynamic CO gas sensing test of SnO₂ at 200°C for carbon monoxide (CO). The sharp loss in sensitivity after testing temperature rose to 250°C, confirmed by impedance spectroscopy, arises from the desorption of chemisorbed oxygen species from the amorphous–crystalline SnO₂ surface. At 200 °C, stable chemisorbed oxygen sustains CO oxidation, yielding the highest and most reproducible sensor response.

The SnO₂ sensor showed strong CO sensitivity at 200°C but a drastic loss of response at 250°C. Impedance spectroscopy confirmed this transition, revealing increased charge-transfer resistance and reduced interfacial capacitance, indicating suppressed surface redox activity. The loss in performance arises from thermal desorption of chemisorbed oxygen ($O^-$, $O_2^-$) species, which are no longer stable at 250 °C—even under oxygen flow—thus eliminating active sites for CO oxidation. TEM analysis revealed mixed amorphous–crystalline phases, where amorphous regions favor oxygen adsorption at lower temperatures but fail to retain these species at higher ones (>250°C). Consequently, 200°C represents the optimal operational window, balancing high chemisorbed oxygen coverage, stable charge transport, and efficient CO–O interaction at the SnO₂ surface.

## 7. Mass Spectroscopy (MS) Analysis

Figure 7 presents MS signals of the CO and $CO_2$ during SnO₂ sensor exposed to carbon monoxide (CO). The left y-axis (black points) shows the CO intensity, while the right y-axis (red points) shows the $CO_2$ intensity. Complementary insights into the surface reactions obtained through operando MS to probe the molecular-level interactions of CO on SnO₂ surface; we conducted simultaneous mass spectrometric analysis in parallel to the O 1s XPS core-level measurements during CO exposure.

At the beginning of the experiment, the CO signal remains low until the CO gas flow starts, after which the CO intensity quickly rises to a steady level. When the CO gas flow begins, the CO signal (black, left axis) quickly rises and remains steady, indicating a continuous CO supply. At the same time, the $CO_2$ signal (red, right axis) spikes sharply because chemisorbed oxygen on the SnO₂ surface reacts with CO to form $CO_2$. As the reaction proceeds, the $CO_2$ signal gradually declines toward baseline as the available chemisorbed oxygen becomes depleted, while the CO signal stays constant. This behavior confirms that chemisorbed oxygen species are responsible for CO oxidation on the SnO₂ surface.

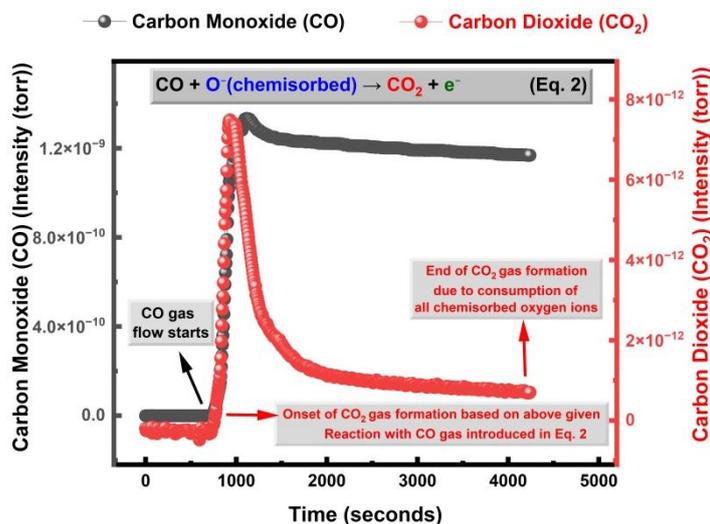

**Figure 7:** Mass spectrometry (MS) profile evolution of carbon monoxide (CO, black) and carbon dioxide ($CO_2$, red) intensities as a function of time during CO exposure of SnO₂ at 200 °C. $CO_2$ formation begins immediately after the onset of CO flow, indicating rapid surface oxidation of CO by chemisorbed oxygen species according to CO + O (chemisorbed) → $CO_2$ + e⁻. The $CO_2$ signal gradually decreases as the chemisorbed oxygen is consumed, marking the end of the surface reaction.

The mass spectrometry (MS) data shows that the CO signal quickly rises to a stable partial pressure of about $1.2 \times 10^{-9}$ torr, indicating a constant CO supply throughout the test. In contrast, the $CO_2$ signal peaks near $8 \times 10^{-12}$ torr, almost two orders of magnitude lower than the CO level. This large difference demonstrates that only a small fraction of the incoming CO is converted to $CO_2$, and that the reaction rate is limited by the finite amount of chemisorbed oxygen ions ($O^-$) on the $SnO_2$ surface. The sharp rise and gradual decay of the $CO_2$ trace reflect the rapid consumption of these reactive oxygen species until they are depleted.

A clear correlation between the detected gaseous product and XPS results (see Figure 3) confirms that carbon dioxide ($CO_2$) is the primary reaction product. Upon the initial introduction of CO, a sharp increase in the $CO_2$ signal was observed; in agreement with the decline in chemisorbed oxygen species indicated by the XPS spectra (see Figure 3). The decrease and then saturation of the $CO_2$ signal in the mass spectrometer aligns precisely with the saturation observed in the sensor response (see Figure 6), suggesting the near-complete consumption of chemisorbed oxygen species. Taken together, these operando measurements provide strong evidence that CO reacts specifically with chemisorbed oxygen species on the $SnO_2$ surface, forming $CO_2$ that subsequently desorbs into the gas phase, as directly detected by mass spectrometry.

## 8. Conclusions

This study clarifies the pivotal role of chemisorbed oxygen ions ($O^-$) in CO sensing over low-temperature-deposited $SnO_2$ thin films exhibiting mixed amorphous–crystalline structure. Transmission electron microscopy (TEM) confirmed the coexistence of amorphous and nanocrystalline regions, which facilitate abundant defect sites and enhance oxygen adsorption. Electrochemical impedance spectroscopy (EIS) identified 200°C as the optimal operating temperature, where charge transfer and surface redox kinetics are most favorable. Combined in situ XPS, mass spectrometry (MS), and sensor analyses verified that CO interacts primarily with chemisorbed oxygen species, generating $CO_2$ and initiating a reversible redox cycle, while lattice oxygen remains inactive. Quantitative surface analysis under 10000 ppm CO and $O_2$ exposures further revealed dynamic modulation of hydroxyl and oxygen species. These findings establish a direct link between microstructural disorder, surface chemistry, and sensing response—offering a mechanistic foundation for designing next-generation, high-performance $SnO_2$-based gas sensors.

## 10. Experimental

**Thin Film Deposition (PEALD):**

$SnO_2$ thin films were deposited by plasma-enhanced atomic layer deposition (PEALD) in a custom stainless-steel reactor (base pressure: $10^{-2}$–$10^{-3}$ mbar; magnetic flux: 2.8 mT). The $Sn(DMP)_4$ precursor was pulsed for 500 ms under 15 sccm $O_2$ and 15 sccm Ar flows, activated by a 13.56 MHz RF plasma (200 W). Films were grown at 60 °C on $Al_2O_3$ substrates including platinum interdigitated electrodes for sensor testing. Substrates were ultrasonically cleaned with isopropanol and water before deposition. **TEM Analysis:** Cross-sectional TEM (FEI Tecnai F20, 200 kV, HAADF) and FIB (FEI Helios G4 CX) were used for microstructural analysis. Final 5 kV ion cleaning minimized beam damage. **In-situ XPS:** Measurements were conducted at the MATLINE beamline (ASTRID2). Spectra were referenced to C 1s = 284.6 eV and Au 3d for Fermi alignment. Peak fitting used Voigt functions with Shirley background subtraction (KolXPD software). **Sensor Tests:** CO sensing was evaluated at 200°C under 1% $O_2/N_2$ balance using a PSS-5 gas control system. Sensor response (Eq. 3) (90% ΔR) was recorded for multiple CO concentrations using with integrated heating and temperature control.

$$\text{Sensor Response (S)} = \left(\frac{R_{Air} - R_{Carbon\ Monoxide\ (CO)}}{R_{Air}}\right) \times 100 \qquad \textbf{(Eq. 3)}$$

## Author Contributions

The data analysis and the original writing of the paper were completed by E.C. The experimental design, planning, and sensor measurements were designed and completed by E.C. and K.S. E.C., Z.L. and K.S. contributed to data analysis and interpretation of literature findings and comparison with the data presented here. E.C. designed the content of the paper. E.C. and Z.L. completed the synchrotron-based measurements. K.S. and E.C. completed the intellectual discussion and impact of the paper. All authors have read and agreed to the published version of the manuscript.

## Funding

Engin Ciftyurek and Klaus Schierbaum thank the European Funds for Regional Development (EFRE-0800672-FunALD) for funding. Engin Ciftyurek acknowledges the utilization of end stations at BESSY II facility under the awarded proposals, with proposal numbers 18106596 and 18106706.

## Acknowledgements

The authors would like to express their sincere gratitude to the staff at Aarhus University for their valuable support during the synchrotron experiments. Author thanks to David Zanders and Anjana Devi for their contribution to the thin-film deposition and TEM analysis. The authors also wish to thank the Institute for Energy and Environmental Technology (IUTA) for their assistance in the design and realization of the sensor experiments.